\documentclass[oribibl]{llncs}

\usepackage[fleqn]{amsmath}

\newcommand{\FMC}{FMC$_{ATINF}$ }
\newcommand{\tup}[1]{\langle #1\rangle}
\newcommand{\la}{\leftarrow}

\begin{document}

\title{Proving Failure of Queries for Definite Logic Programs Using
  XSB-Prolog}
\author{Nikolay Pelov and Maurice Bruynooghe}

\institute{
Departement Computerwetenschappen \\
Katholieke Universiteit Leuven \\
Celestijnenlaan 200A \\
B-3001 Heverlee, Belgium \\
E-mail: \email{\{pelov,maurice\}@cs.kuleuven.ac.be}}

\maketitle

\begin{abstract}
Proving failure of queries for definite logic programs can be done by
constructing a finite model of the program in which the query is
false. A general purpose model generator for first order logic can be
used for this. A recent paper presented at PLILP98 shows how the 
peculiarities of definite programs can be exploited to obtain a better
solution. There a procedure is described which combines abduction with
tabulation and uses a meta-interpreter for heuristic control of the
search. The current paper shows how similar results can be obtained
by direct execution under the standard tabulation of the XSB-Prolog
system. The loss of control is compensated for by better intelligent
backtracking and more accurate failure analysis.
\end{abstract}

\section{Introduction}

In \cite{fq:mb-plip98} methods are studied for proving that a query
for a definite logic program fails. The general idea underlying all
methods is the generation of a finite model of the definite program in
which the query is false. However the approach developed in
\cite{fq:mb-plip98} is quite different from that used in general
purpose model generators for first order logic such as FINDER
\cite{mg:finder}, SEM \cite{mg:sem}, and \FMC \cite{mg:peltier-98}.
Whereas the latter systems search for a model in the space of
interpretations, the former searches in the smaller space of
pre-interpretations and applies a top-down proof procedure using
tabulation to verify whether the query is false in the least model of
the Horn theory based on the candidate pre-interpretation. Experiments
in \cite{fq:mb-jflp}, an extended version of \cite{fq:mb-plip98}, show
that the abductive procedure of \cite{fq:mb-plip98} extended with
intelligent backtracking \cite{ib:mb-81} outperforms FINDER and \FMC
on problems where there are a large number of different
interpretations for a given pre-interpretation. The difference is not
only in the number of backtracks, but also, for some problems, in
time, and this notwithstanding the former is implemented as a
straightforward meta-interpreter in Prolog while the latter are
sophisticated implementations in a more low level language.

The current paper describes how the meta-interpreter can be replaced
by a more direct implementation in XSB-Prolog \cite{xsb,slg} which
relies on the XSB system to perform the tabulation. This is not a
straightforward task because of the intelligent backtracking and
because the meta-interpreter does not follow the standard depth-first
left-to-right search strategy but uses heuristics to direct the search
towards early failures and selects the pre-interpretation on the fly,
as components are needed by the proof procedure. To exploit the
tabling system underlying XSB, one has to stick to the depth-first 
left-to-right execution order and one should not modify the program by
creating new components of the pre-interpretation while evaluating a
call to a tabled predicate.

The random selection of an initial pre-interpretation, combined with
the loss of control over the search results in a system which has to
explore a substantially larger part of the search space than the
original system. The paper introduces two innovations to compensate for
this. Firstly, it uses a variant of intelligent backtracking which is
much less dependent on the random initial order of the choice points.
Secondly, it introduces a more accurate failure analysis, so that
smaller conflict sets are obtained and that the intelligent
backtracking selects its targets with more accuracy.

The motivation for this research is in the world of planning. 
Planners are typically programs which search in an infinite space of
candidate plans for a plan satisfying all requirements. The planner
searches forever (until some resource is exhausted) when no candidate
plan satisfies all requirements. Hence it is useful to have methods to
show that the problem has no solution. It turns out that our approach
outperforms first order model generators on planning problems.

In the next section we recall some basic notions about semantic of
definite logic programs. In Section~\ref{sec:method} we describe our
approach in more detail and then in Section~\ref{sec:exper} we show
the results of testing our system on different problems. 
The comparison not only includes the model generator FINDER
\cite{mg:finder} as in \cite{fq:mb-plip98}, and \FMC as in
\cite{fq:mb-jflp} but also SEM \cite{mg:sem}.

\section{Preliminaries} \label{sec:prel}

Now we will recall some basic definitions about semantics of definite
programs. Most of them are taken from \cite{lloyd}.

A {\it pre-interpretation} $J$ of a program $P$ consists of domain
$D=\{d_1,\ldots,d_m\}$\footnote{We will consider only domains with
finite size.} and for each $n$-ary function symbol $f$ in $P$ a
mapping $f_J$ from $D^n$ to $D$. 
Following the literature on model generators, a term of the form
$f(d_1,\ldots,d_n)$ where $d_1,\ldots,d_n\in D$ is called a {\it cell}.
Given a program $P$ and domain size $m$, the set of all cells is fixed. 
A pair $\tup{c,v}$ where $c$ is a cell and $v\in D$ is the mapping of
that cell is called a {\it component} and $v$ the {\it value} of the
component.  
A set of components defines a pre-interpretation if there is exactly
one component $\tup{c,v}$ for each cell.

A {\it variable assignment} $V$ wrt.\ expression $E$ and
pre-interpretation $J$ consists of an assignment of an element in the
domain $D$ for each variable in $E$. A {\it term assignment} wrt.\ $J$
and $V$ is defined as follows: each variable is given its assignment
according to $V$; each constant is given its assignment according to
$J$;  if $d_1,\ldots,d_n$ are the term assignments of $t_1,\ldots,t_n$
then the assignment of $f(t_1,\ldots,t_n)$ is the value of the cell
$f(d_1,\ldots,d_n)$. 

An interpretation $I$ based on a pre-interpretation $J$ consists of a
mapping $p_I$ from $D^n$ to $\{false, true\}$ for every $n$-ary
predicate $p$ in $P$. An interpretation $I$ is often defined as the
set of atoms $p(d_1,\ldots,d_n)$ for which $p(d_1,\ldots,d_n)$ is
mapped to true. An interpretation $M$ is a model of a program $P$ 
iff all clauses in $P$ are true in $M$. For a definite program, the
intersection of two models is also a model hence a definite program
always has a unique least model.
As a consequence, if a conjunction of atoms is false in some model
then it is also false in the least model of a definite program.

Throughout the paper we will use the following simple example about
even and odd numbers to show the different concepts and program
transformations.

\begin{verbatim}
even(zero).
even(s(X)) :- odd(X).
odd(s(X)) :- even(X).
\end{verbatim}

Consider a query {\tt ?- even(X),odd(X)}. For simplicity of the
presentation we will add to the program the definite clause 

\begin{verbatim}
even_odd :- even(X),odd(X).
\end{verbatim}

and consider the query {\tt ?- even\_odd}. 
It cannot succeed as {\tt ?- even\_odd} is not a logical
consequence of the program. The SLD proof procedure does not
terminate. This is still the case when extended with tabulation as in
XSB-Prolog.

We choose a domain with two elements $D=\{0,1\}$ and consider 
the pre-interpretation $J=\{zero_J=0, s_J(0)=1, s_J(1)=0\}$.
The least model of the definite program is $\{even(0), odd(1)\}$
and the atom {\tt even\_odd} is false in this model.

\section{The Method} \label{sec:method}

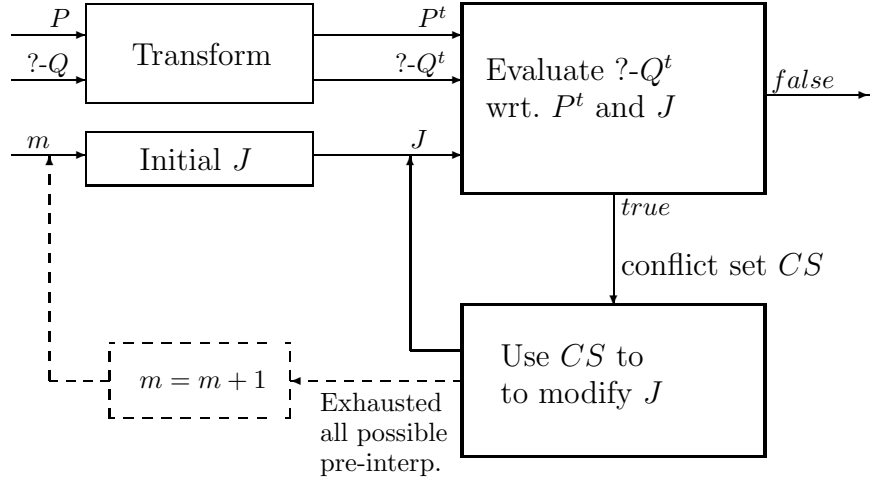
\begin{figure}[htbp]
  \begin{center}
\setlength{\unitlength}{4144sp}%
\begingroup\makeatletter\ifx\SetFigFont\undefined%
\gdef\SetFigFont#1#2#3#4#5{%
  \reset@font\fontsize{#1}{#2pt}%
  \fontfamily{#3}\fontseries{#4}\fontshape{#5}%
  \selectfont}%
\fi\endgroup%
\begin{picture}(5154,2852)(214,-2441)
\thinlines
\put(3826,-736){\vector( 0,-1){675}}
\put(226,209){\vector( 1, 0){450}}
\put(4726,-151){\vector( 1, 0){630}}
\multiput(2926,-1861)(-108.94737,0.00000){10}{\line(-1, 0){ 54.474}}
\put(1891,-1861){\vector(-1, 0){0}}
\put(811,-2086){\dashbox{57}(1080,450){}}
\put(676,-691){\framebox(1350,315){}}
\put(676,-196){\framebox(1350,585){}}
\put(2026,-61){\vector( 1, 0){900}}
\put(226,-61){\vector( 1, 0){450}}
\put(2026,-511){\vector( 1, 0){900}}
\put(226,-511){\vector( 1, 0){450}}
\put(2026,209){\vector( 1, 0){900}}
\thicklines
\put(2926,-736){\framebox(1800,1125){}}
\put(2926,-2311){\framebox(1800,900){}}
\thinlines
\multiput(811,-1861)(-102.85714,0.00000){4}{\line(-1, 0){ 51.429}}
\multiput(451,-1861)(0.00000,117.39130){12}{\line( 0, 1){ 58.696}}
\put(451,-511){\vector( 0, 1){0}}
\put(2926,-1681){\line(-1, 0){315}}
\put(2611,-1681){\vector( 0, 1){1170}}
\put(4771,-106){\makebox(0,0)[lb]{\smash{\SetFigFont{10}{12.0}{\familydefault}{\mddefault}{\updefault}$false$}}}
\put(3871,-1231){\makebox(0,0)[lb]{\smash{\SetFigFont{12}{14.4}{\familydefault}{\mddefault}{\updefault}conflict set $CS$}}}
\put(991,-1906){\makebox(0,0)[lb]{\smash{\SetFigFont{10}{12.0}{\familydefault}{\mddefault}{\updefault}$m=m+1$}}}
\put(946, 29){\makebox(0,0)[lb]{\smash{\SetFigFont{12}{14.4}{\familydefault}{\mddefault}{\updefault}Transform}}}
\put(991,-601){\makebox(0,0)[lb]{\smash{\SetFigFont{12}{14.4}{\familydefault}{\mddefault}{\updefault}Initial $J$}}}
\put(2071,-2221){\makebox(0,0)[lb]{\smash{\SetFigFont{10}{12.0}{\familydefault}{\mddefault}{\updefault}all possible}}}
\put(2611,-466){\makebox(0,0)[lb]{\smash{\SetFigFont{10}{12.0}{\familydefault}{\mddefault}{\updefault}$J$}}}
\put(316,-466){\makebox(0,0)[lb]{\smash{\SetFigFont{10}{12.0}{\familydefault}{\mddefault}{\updefault}$m$}}}
\put(3871,-871){\makebox(0,0)[lb]{\smash{\SetFigFont{10}{12.0}{\familydefault}{\mddefault}{\updefault}$true$}}}
\put(2071,-2041){\makebox(0,0)[lb]{\smash{\SetFigFont{10}{12.0}{\familydefault}{\mddefault}{\updefault}Exhausted}}}
\put(3061,-286){\makebox(0,0)[lb]{\smash{\SetFigFont{12}{14.4}{\familydefault}{\mddefault}{\updefault}wrt. $P^t$ and $J$}}}
\put(316,-16){\makebox(0,0)[lb]{\smash{\SetFigFont{10}{12.0}{\familydefault}{\mddefault}{\updefault}?-$Q$}}}
\put(3061,-61){\makebox(0,0)[lb]{\smash{\SetFigFont{12}{14.4}{\familydefault}{\mddefault}{\updefault}Evaluate ?-$Q^t$}}}
\put(2521,-16){\makebox(0,0)[lb]{\smash{\SetFigFont{10}{12.0}{\familydefault}{\mddefault}{\updefault}?-$Q^t$}}}
\put(451,254){\makebox(0,0)[lb]{\smash{\SetFigFont{10}{12.0}{\familydefault}{\mddefault}{\updefault}$P$}}}
\put(2656,254){\makebox(0,0)[lb]{\smash{\SetFigFont{10}{12.0}{\familydefault}{\mddefault}{\updefault}$P^t$}}}
\put(3151,-1771){\makebox(0,0)[lb]{\smash{\SetFigFont{12}{14.4}{\familydefault}{\mddefault}{\updefault}Use $CS$ to}}}
\put(3151,-1996){\makebox(0,0)[lb]{\smash{\SetFigFont{12}{14.4}{\familydefault}{\mddefault}{\updefault}to modify $J$}}}
\put(2071,-2401){\makebox(0,0)[lb]{\smash{\SetFigFont{10}{12.0}{\familydefault}{\mddefault}{\updefault}pre-interp.}}}
\end{picture}

    \caption{System architecture}
    \label{fig:arch}
  \end{center}
\end{figure}

Figure~\ref{fig:arch} shows the general architecture of the system.
The input consists of a definite program $P$, a query ?-$Q$ and
domain size $m$. First the program and the query are transformed to
$P^t$ and ?-$Q^t$. The transformation replaces all functional symbols
with calls to predicates defining the components of the
pre-interpretation and allows the program to 
collect the components which were used during the evaluation of the query.
Also an initial pre-interpretation $J$ is constructed for the given domain
size $m$. Then the query ?-$Q^t$ is evaluated wrt.\ the program $P^t$ and
the current pre-interpretation $J$. If the query succeeds then it
also returns a set of components $CS$ which are necessary for the
success of the proof. 
Then, based on $CS$, the pre-interpretation is modified and the query
is run again. 
If we have exhausted all possible pre-interpretations for the given
domain size then we can eventually increase it and run the system again.
If the query ?-$Q^t$ fails then $Q^t$ is false in the least model
based on the pre-interpretation $J$ and we can conclude that the
original query ?-$Q$ cannot succeed.

\subsection{Basic Transformation} \label{sec:basic}

To evaluate the query in the least model based on a pre-interpretation
$J$, we use a variant of the abstract compilation approach to program
analysis used by Codish and Demoen in \cite{cd}. The
pre-interpretation $J$ of a $n$-ary function $f$ is represented by a
set of facts $p_f(d_1,\ldots,d_n,v)$; one fact for each cell
$f(d_1,\ldots,d_n)$.  In the source program, non variable terms are
represented by their pre-interpretation. This is achieved by replacing
a term $f(t_1,\ldots,t_n)$ by a fresh variable $X$ and introducing a
call $p_f(t_1,\ldots,t_n,X)$. This transformation is repeated for the
non variable terms in $t_1,\ldots,t_n$ until all functions are
eliminated. Codish and Demoen evaluate the resulting DATALOG program
bottom up, obtaining the least model which expresses declarative
properties of the program. In \cite{fq:mb-plip98}, one also transforms
the query and using a top-down procedure with tabulation checks
whether it fails.  Experience showed that one typically ends up with
computing the whole model of the predicates reachable from the query.
So the meta-interpreter used there tables only the most general call
for each predicate. As we want direct execution under XSB, our
transformation has to take care that a program predicate is only
called with all variables free and different, so that XSB tables only
the most general call. To achieve this, a predicate $p_f(\ldots)$
which is added to compute a term $t$ in a call is inserted after the
call and a predicate which is added to compute a term in the head is
inserted at the end of the clause. Finally, when a call to a program
predicate contains a variable $X$ which already occurs to the left of
its position in the clause, then it is replaced by a fresh variable
$Y$ and an equality $X=Y$ is inserted after the call. The calls to the
pre-interpretation are not tabled, and a call $p_f(g(\ldots),\ldots)$
is transformed in $p_g(\ldots,X),p_f(X,\ldots)$. This gives less
branching than when $p_g(\ldots)$ is added after $p_f(\ldots)$. For
our example this gives the following code:

\begin{verbatim}
even(X) :- p_zero(X).
even(Y) :- odd(X),p_s(X,Y).
odd(Y) :- even(X),p_s(X,Y).
even_odd :- even(X),odd(X1),X1=X.

p_zero(0).
p_s(0,1).
p_s(1,0).
\end{verbatim}

In \cite{fq:mb-plip98}, values are assigned to the cells of the
pre-interpretation in an abductive way, as needed by the heuristic
search for a proof of the query. When a proof is found, standard
backtracking occurs: the last assigned value is modified. To have
direct execution under XSB, the pre-interpretation has to be fixed in
advance. Obviously, it is not feasible to enumerate all possible
pre-interpretations until one is found for which the query fails. The
search has to be guided by the proof found so far. Failure analysis
and intelligent backtracking have to be incorporated to obtain a
usable system.

\subsection{Failure Analysis} \label{sec:fa}

\subsubsection{Elementary Failure Analysis.}

As the goal is to find a pre-interpretation for which the query fails,
failure occurs when the query succeeds. In the more general setting of
first order model generation, failure occurs when some formula gets
the wrong truth value. 
The FINDER and \FMC systems keep track of which cells are used in
evaluating a formula and when the formula receives the wrong truth
value, the set of cells used in evaluating it is used to direct the
backtracking. In \cite{fq:mb-jflp} the meta-interpreter is extended
with such a failure analysis and intelligent backtracking is used to
guide the search. This substantially improved the performance of the
system. Incorporating these features in the current approach which
relies on direct execution with XSB of the transformed query, requires
special care. First let us formalize the notion of conflict set
(refutation in first order model generators
\cite{mg:peltier-98,mg:finder}). 

\begin{definition}[Conflict set]
A {\em conflict set} $CS$ of a definite program $P$ and query $Q$ is a
finite set of components such that for any pre-interpretation $J$ for
which $CS\subseteq J$ follows that $Q$ is true in any model of $P$
based on $J$.
\end{definition}

The idea is that any pre-interpretation $J$ which has the same values
for all components from the conflict set $CS$ can not be extended to an
interpretation in which the query fails. Hence any candidate
pre-interpretation must differ from $CS$ in the value of at least one
component. 
Exploiting conflict sets requires first to compute them.
This can be done by adding to the program predicates an extra argument
which is used to collect the components used for solving a call to
this predicate. For example a
call {\tt even(X)} is replaced by {\tt even(X,CS)} and the answer 
{\tt even(0)} becomes {\tt even(0,[p\_zero(0)])}.
However there is a potential problem. Also
{\tt even(0,[p\_zero(0),p\_s(0,1),p\_s(1,0)])} is an
answer. Previously, the tabling system did not recognize it as a new
answer and did not use it to solve calls to {\tt even/1}. 
But as the value of the added second argument differs from that in the
first answer, XSB will also use it to solve calls to $even/2$ and it
will obtain a third answer. Fortunately, if the list of used
components is reduced to some canonical form, then the third answer
will be identical to the second and the evaluation will terminate.
However, this repetition of answers with different lists of components
can substantially increase the cost of the query evaluation.
Fortunately the XSB system has built-in predicates to inspect and
modify the tables so we can control this behavior.
The idea is to replace a clause

\begin{verbatim}
   p(X,CS) :- Body.
\end{verbatim}

\noindent with a clause

\begin{verbatim}
   p(X,CS) :- Body,check_return(p(X,CS)).
\end{verbatim}

When the body of the clause succeeds, XSB will process the answer
$p(X,CS)$ (add it to the table for the call to $p/2$ if it is new).
Remember, that as the transformed program makes only most general
calls there is only one table associated with each predicate. Using
the built-ins, the predicate $check\_return/1$ looks up the previous
answers in the table for $p/2$ and compares them with the candidate
answer $p(X,CS)$. If there is no other answer with the same $X$ then
$check\_return/1$ and thus $p/2$ simply succeed. The interesting case
is when the table already holds an answer $p(X,CS_{old})$ with a
different conflict set $CS_{old}$ (if $CS_{old}=CS$ then XSB will
recognize it is a duplicate answer).  Then several strategies are
possible for {\tt check\_return/1}:
\begin{itemize}
\item The simplest approach is to let {\tt check\_return/1} fail when
  the table already holds an answer with the same $X$.
\item An alternative approach is to check whether the new conflict set
  $CS$ is ``better'' than $CS_{old}$. Then the old answer is removed
  from the table and {\tt check\_return/1} succeeds. Otherwise 
  {\tt check\_return/1} fails.
\item Finally, but more expensive for the overall query evaluation, one
  could allow several answers, only rejecting/removing redundant ones
  ($p(X,CS_1)$ is redundant wrt. $p(X,CS_2)$ if $CS_1\supseteq CS_2$). 
\end{itemize}

\subsubsection{Advanced Failure Analysis.}

A conflict set can be called minimal if it has no subset which is a
conflict set. Obviously it is not feasible to compute minimal conflict
sets. However, simply collecting the components used in a proof can be
a large overestimation. For example, in our planning problems, a three
argument predicate is used: one argument is the initial state, one
argument is the final state and one argument is the description of the
derived plan. The pre-interpretation of the terms representing the plan
is completely irrelevant for the failure of the query. However the
components used to compute it will be part of the conflict set.

To see how to refine our failure analysis, let us reconsider how
answers are obtained. Using a slightly different notation, the base
case of the $even/1$ predicate can be written as:

\begin{verbatim}
even(X) :- X=0_J.
\end{verbatim}

This represents the basic answer, parameterized by the pre-interpretation
$J$. Now consider the definition of the $odd/1$ predicate:

\begin{verbatim}
odd(X) :- even(Y),X=s_J(Y).
\end{verbatim}

An answer of $odd/1$ is obtained by performing resolution with the
basic answer for $even/1$, yielding:

\begin{verbatim}
odd(X) :- Y=X1,X1=0_J,X=s_J(Y).
\end{verbatim}

This can be generalized, answers for a predicate $p/n$ are of the
form:
\[p(X_1,\ldots,X_n)\la X_1=t_{1_J},\ldots,X_N=t_{n_J}, Eqs\]
with $Eqs$ a set of equations involving $X_1,\ldots,X_n$ and some
local variables $Y_1,\ldots,Y_n$. Under the elementary failure analysis
the answer is $p(t_{1_J},\ldots,t_{n_J})$ and the associated conflict
set is the set of components used in computing
$t_{1_J},\ldots,t_{n_J}$ and the terms of $Eqs$.

The basis for the advanced failure analysis is the observation that
the answer clauses can be simplified while preserving the solution
they represent.
Terms form equivalence classes under a pre-interpretations.
Members of the equivalence class can be represented by the domain
element which is their pre-interpretation and equalities between terms
modulo equivalence class can be simplified using three of the four
Martelli-Montanari simplification rules:
\begin{itemize}
\item $p(t_{1_J},\ldots,t_{n_J})\la X=X,Eqs$ is equivalent to \\
      $p(t_{1_J},\ldots,t_{n_J})\la Eqs$ (remove)
\item $p(t_{1_J},\ldots,t_{n_J})\la t_J=X,Eqs$ is equivalent to \\
      $p(t_{1_J},\ldots,t_{n_J})\la X=t_J,Eqs$ (switch)
\item $p(t_{1_J},\ldots,t_{n_J})\la X=t_J,Eqs$ is equivalent to \\
      $p(t_{1_J},\ldots,t_{n_J})\{X/t_J\}\la Eqs\{X/t_J\}$ (substitute)
\end{itemize}

Note that $f_J(t_{1_J},\ldots,t_{n_J})=g_J(s_{1_J},\ldots,s_{m_J}),
Eqs$ is not equivalent to $false$ and that
$f_J(t_{1_J},\ldots,t_{n_J})=f_J(s_{1_J},\ldots,s_{n_J}), Eqs$ is not
equivalent to $t_{1_J}=s_{1_J},\ldots,t_{n_J}=s_{n_J}, Eqs$, hence
peel is not allowed.

So an answer can be simplified to a form 
\[p(t_{1_J},\ldots,t_{n_J})\la Eqs\]
where $Eqs$ contains equations between non variable terms and
some of the $t_{i_J}$ in the head can be variables.
The pre-interpretations in the terms of $Eqs$ decide whether $Eqs$
is interpreted as true or false, hence the components used in
interpreting the terms in $Eqs$ form the real conflict set of the
answer. However also the components used to interpret the terms
$t_{i_J}$ of the head are important.
When the answer is used to solve a call, they become part of new
equations. 
Hence, with each variable we should associate a set holding the
components used in evaluating the term the variable is bound to and
with each answer we should associate the ``real'' conflict set.
Moreover, the execution of the equalities $X=Y$ has to be monitored. 
When one of $X$ or $Y$ is free then unification can be performed,
otherwise if $X$ and $Y$ have the same interpretation then
the sets of components associated with $X$ and $Y$ have to be added
to the conflict set of the answer (as before the equality fails when
$X$ and $Y$ have a different interpretation). Note that our
transformation is such that calls have fresh variables as arguments,
so the equality between an argument of a call and an argument of an
answer always involves a free variable and is correctly handled by
standard unification.
A final point is that the body of the compiled clause has to be
carefully ordered: equalities on predicate calls involving a variable
$X$ should precede the interpretation of a term containing $X$,
e.g. $p(X), Y=f_J(X)$ is a correct ordering: first the call $p/1$
binds $X$ to a domain element and also returns the set of components
$CS_X$ used in computing that domain element.
Then $Y$ is bound to a domain element and the set of components used in
computing it is $\{f_J(X)\}\cup CS_X$.
Taking the above into account, the code for our example is as follows:

\begin{verbatim}
even(X,[]) :- comp(p_zero,[],X), check_return(even(X,[])).
even(X,CS) :- odd(Y,CS),comp(p_s,[Y],X), check_return(even(X,CS)).
odd(X,CS) :- even(Y,CS),comp(p_s,[Y],X), check_return(odd(X,CS)).
even_odd(CS) :-
     even(X,EvenCS),odd(Y,OddCS),
     merge(EvenCS,OddCS,CS1),unify(X,Y,CS1,CS),
     check_return(even_odd(CS)).
\end{verbatim}

Calls to the pre-interpretation are made through an intermediate
predicate {\tt comp/3} defined below. The call to 
{\tt combine\_arg\_cs/3} collects the conflict sets associated with the
ground arguments of the function to be interpreted (none if the
argument is a free variable) in {\tt ArgsCS} and {\tt merge/3} extends
{\tt ArgsCS} with {\tt Comp}, the consulted component of the
pre-interpretation, to obtain the final conflict set {\tt ResCS}.

\begin{verbatim}
comp(F,Args,R-ResCS) :-
    combine_arg_cs(Args,RealArgs,ArgsCS),
    append([F|RealArgs],[R],C),Comp =.. C,
    call(Comp),
    merge([Comp],ArgsCS,ResCS).

combine_arg_cs([],[],[]).
combine_arg_cs([A-[]|T],[A|T1],RestCS) :- !,
    combine_arg_cs(T,T1,RestCS).
combine_arg_cs([A-ACS|T],[A|T1],OutCS) :-
    combine_arg_cs(T,T1,RestCS),
    merge(ACS,RestCS,OutCS).
\end{verbatim}

The {\tt merge/3} predicate makes the union of two sets (represented
as lists) and places the result in a canonical form and {\tt unify/4}
is used to monitor the unification process and can be defined by the
following Prolog code:

\begin{verbatim}
unify(X,Y,S,S) :- (var(X);var(Y)), !, X=Y.
unify(X-Sx,X-Sy,Sin,Sout) :- merge(Sx,Sy,S), merge(S,Sin,Sout).
\end{verbatim}

The first two arguments are the terms to be unified, the third is the
current conflict set of the clause and the last argument is the new
conflict set of the clause.  The first clause handles the case that
one is a free variable: unification is performed and the conflict set
of the clause remains the same. The second clause handles the case
that both arguments $X$ and $Y$ are bound to the same domain element.
The set of components used in evaluating the first argument $(Sx)$ and
in evaluating the second argument $(Sy)$ are added to $Sin$ yielding
$Sout$.

\subsection{Intelligent Backtracking}

Under standard backtracking, candidate pre-interpretations are
enumerated according to some fixed total ordering $c_1,c_2,\ldots,c_n$
of the cells. 
When some partial solution $c_1=d^1_1,c_2=d^1_2,\ldots,c_m=d^1_m$
is rejected then the value assignment $d^1_m$ for the last cell $c_n$
is modified. If no other value is left, then $c_{m-1}$ is modified
(and all domain elements become again available for $c_m$).
The simplest use of conflict sets is based on the observation that no
extension of the conflict set can be a solution, so the last element
according to the total order over the cells of the conflict set is
selected and the assignment to this cell is modified. However also
secondary conflict sets can be derived \cite{ib:mb-81}.
Assume, due to different conflicts, all values for some cell $c_n$
have been rejected. With $\{c_{i,1},\ldots,c_{i,k_i},c_n\}$ the
conflict set which led the rejection of $d_i$ we can formalize the
knowledge in the conflict sets as:
\begin{eqnarray*}
 & & c_{1,1}=d_{1,1}\wedge\ldots\wedge c_{1,k_1}=d_{1,k_1}\wedge c_n=d_1 
 \rightarrow false \\
 & & \vdots \\
 & & c_{m,1}=d_{m,1}\wedge\ldots\wedge c_{m,k_m}=d_{m,k_m}\wedge c_n=d_m
 \rightarrow false.
\end{eqnarray*}
As we have that cell $c_n$ must be assigned some domain element, we
have $c_n=d_1 \vee \ldots \vee c_n=d_m$.
Applying hyper-resolution \cite{robinson}, one can infer
\begin{eqnarray*}
& & c_{1,1}=d_{1,1}\wedge\ldots\wedge c_{1,k_1}=d_{1,k_1}\wedge \\
& & \vdots \\
& & c_{m,1}=d_{m,1}\wedge\ldots\wedge c_{m,k_m}=d_{m,k_m}\rightarrow false
\end{eqnarray*}
which says that 
$\{c_{1,1},\ldots, c_{1,k_1},\ldots,c_{m,1},\ldots,c_{m,k_m}\}$
is also a conflict set.

At the implementation level, an accumulated conflict set is associated
with each cell and initialized as empty. When a conflict
$\{c_1,\ldots,c_{n-1},c_n\}$ is derived with $c_n$ its last cell, then 
$\{c_1,\ldots,c_{n-1}\}$ is added to the accumulated conflict set of
$c_n$. Once all assignments to a cell are exhausted, its associated
conflict set holds the secondary conflict which can be used
to direct further backtracking. This is the approach taken in 
\cite{fq:mb-jflp} where it worked quite well, as the initial order
was carefully chosen. In the current implementation, where the
initial order over the cells is random, the system had to do much
more search before finding a solution. Hence we adopted a variant of
intelligent backtracking mentioned in \cite{ib:mb-81} which leaves the
cells unordered until they participate in a conflict. Under this
approach, cells are split over two sets, a set with a total order
(initially empty) and a set which is unordered. When a conflict is
found, the cells from it which are in the unordered set
(if any) are moved to the end of the ordered set. Then the last cell
of the conflict set is chosen as target of the backtracking. Cells
which are after the target in the total order return to the unordered
set. This approach resulted in substantially better results.

\subsection{Dealing with Equational Problems} \label{sec:eq}

There exists many problems which contain only one predicate, the
equality predicate $eq/2$. They consist of a number of facts
$eq(t_{i_1},t_{i_2})\la$ for $i = 1,\ldots,m$ and a number of denials
$\la eq(s_{j_1},s_{j_2})$ for $j=1,\ldots,n$.
To solve such problems, one has to add to the program the 
axioms for the equality theory for reflexivity, symmetry, transitivity
and function substitution, the latter consists of an axiom 
\[f(X_1,\ldots,X_n) = f(Y_1,\ldots,Y_n) \la 
   X_1 = Y_1\wedge\ldots\wedge X_n = Y_n. \]
for each functor $f/n$. The least model of the
standard equality theory is the identity relation over the domain of
the interpretation, hence the search space can be reduced by
restricting the interpretation of $eq/2$ to the identity relation.

In the abductive system of \cite{fq:mb-jflp}, this is achieved by
initializing the interpretation of $eq/2$ as identity, and removing
the standard equality theory (only the problem specific facts
and denials remain). Backtracking is initiated as soon as either one
of the denials $eq(s_{j_1},s_{j_2})$ evaluates to true or one of the
facts $eq(t_{i_1},t_{i_2})$ results in an answer which is not in the
identity relation.

With direct execution under XSB, a slightly different approach is
required. Unification reduces to the identity relation, hence after
compiling the terms, the call to $eq/2$ can be done by unifying the
compiled terms. However, the problem is that all facts and denials
need to be activated. Therefore a new predicate $p/0$ is introduced
and defined as follows:
\begin{alignat*}{2}
& p\la \neg eq(t_{i_1},t_{i_2}). &\quad & i = 1,\ldots,m\\
& p\la eq(s_{j_1},s_{j_2}).      && j = 1,\ldots,n
\end{alignat*}
Proving failure of the query $\la p$ yields the desired
pre-interpretation. Indeed $p$ is equivalent to
\[
p\la \bigvee_{1\leq i\leq m}\exists\; \neg eq(t_{i_1},t_{i_2}) \vee
     \bigvee_{1\leq j\leq n}\exists\; eq(s_{j_1},s_{j_2}).
\]
Hence $p$ fails if the right-hand side is true, i.e. if
\[
 \bigwedge_{1\leq i\leq m}\forall\; eq(t_{i_1},t_{i_2}) \wedge
 \bigwedge_{1\leq j\leq n}\forall\; \neg eq(s_{j_1},s_{j_2})
\]
is true. $\forall\, eq(t_{i_1},t_{i_2})$ is equivalent with the fact
$eq(t_{i_1},t_{i_2})$ and $\forall\, \neg eq(s_{j_1},s_{j_2})$ is
equivalent to the denial $\la eq(s_{j_1},s_{j_2})$. Thus $p$ fails if
the conjunction of the original facts and denials is true under the
chosen pre-interpretation. Compilation of terms is as described in
Section~\ref{sec:basic}, i.e. a call $eq(s_{j_1},s_{j_2})$ is replaced
by a call $X_{j_1}=X_{j_2}$ preceded by the code computing the
pre-interpretation of $s_{j_1}$ and $s_{j_2}$. A call $\neg
eq(t_{i_1},t_{i_2})$ is handled in a similar way; the built-in
$\backslash\!\!=$ (not unifiable) can be used instead of not equal.
However, special care is required to ensure the arguments
are ground in case $t_{i_1}$ or $t_{i_2}$ is a variable. Whereas the
compilation 
leaves such variables intact, here it has to be mapped (the mapping
introduces a backtrack point) to a domain element.

Similarly as in Section~\ref{sec:fa}, conflict sets can be associated
with terms for the task of advanced failure analysis.  Hence a call
$\neg eq(t_{i_1},t_{i_2})$ is transformed in the sequence
$interpret(t_{i_1},X_{i_1}),interpret(t_{i_2},X_{i_2}),
disunify(X_{i_1},X_{i_2},S_{in},S_{out})$ where $interpret/2$ is an
abbreviation for the sequence of calls computing the
pre-interpretation of the term and the associated conflict set and
$disunify/4$ is defined as
\begin{verbatim}
disunify(X-Sx,Y-Sy,Sin,Sout) :- 
    X\=Y,merge(Sx,Sy,S), merge(S,Sin,Sout).
\end{verbatim}

\section{Experiments} \label{sec:exper}

\subsection{The Problems}

We tested our system with a large number of different problems.
Below we give a short description for each one of them and for some of
them the source code is given in Appendix~\ref{app:code}.

\subsubsection{List Manipulation.}
The {\tt appendlast} problem uses the standard definition of the
predicates {\tt append} and {\tt last} and the following query:

\begin{verbatim}
appendlast :- append(X, [a], Xs),last(Xs, b).
\end{verbatim}

\noindent
The {\tt reverselast} problem is similar to the {\tt appendlast} problem
but uses the version of the predicate {\tt reverse} with accumulator:

\begin{verbatim}
reverselast:- reverse(L, R, [a]), last(R, b).
\end{verbatim}

\noindent
The {\tt nreverselast} problem uses the ``naive'' definition of 
{\tt reverse}: 

\begin{verbatim}
nreverselast :- reverse([a|X], R), last(R, b).
\end{verbatim}

\subsubsection{Multisets.}
The {\tt multiset?o} are programs to check the
equivalence of two multisets using a binary operator ``o'' to
represent them. {\tt multiset3o} is a problem which has a solution,
thus failure cannot be proven for it.

\subsubsection{Planning in the Blocks-World.}

These are simple problems for planning in the blocks-world.
The theory for the {\tt blockpair} problems has, besides the usual
actions of the blocks-world, an action to add or remove a pair of
blocks. In the {\tt blockzero} problems, the extra action is to create
a new block named $s(X)$ on top of a clear block $X$.

The queries ending in ``o'' use multisets based on the function
{\tt o/2} and those ending in ``l'' use a standard list representation.
Those problems which have the number 2 in their name do not collect
the plan and those having 3 store the plan in the second argument.
{\tt blockzero2ls}\footnote{corresponds to {\tt blocksol} in
  \cite{fq:mb-plip98} and \cite{fq:mb-jflp}}
is a problem which has a solution.

\subsubsection{TPTP-Problems.}

The rest of the examples are taken from the TPTP problem library 
\cite{tptp}. In Table~\ref{tab:prop} in brackets are given the TPTP
names for each one of them.
All these problems are equational problems and are transformed in the
way described in Section~\ref{sec:eq}.

\smallskip\noindent
The {\tt tba} problem is to prove an independence of one axiom for
ternary boolean algebra. 

\smallskip\noindent
The {\tt grp} problem is to prove that some axiom is not a single
axiom for group theory.

\smallskip\noindent
The {\tt cl3} problem is from the domain of combinatory logic and the
goal is to find a set of combinators which satisfy
axioms $S$ and $W$ and do not satisfy the weak fixed point property.

\medskip\noindent
Table~\ref{tab:prop} gives some details about the properties of the 
problems. The column {\em \#pred} shows the number of predicates.
The column {\em size~dom}\ gives the domain size for which the query
has been evaluated (which is, for the failing queries, the minimum
domain size for which a model proving failure exists).
The column {\em size~pre}\ gives the number of cells in
the pre-interpretation and the next column {\em \#pre}\ gives the
number of all possible pre-interpretations for the given domain size.
The column {\em size~int}\ gives the number of atoms to be assigned a
truth value in an interpretation and the last column {\em \#int/pre}\ 
gives the number of different interpretations for a fixed
pre-interpretation. For the TPTP problems this value is 1 because
they have only one predicate for which the interpretation is known to
be identity.

\begin{table}
\caption{Example properties} \label{tab:prop}
\begin{center}
\begin{tabular}{|l|r|r|r|r|r|r|r|} 
\hline
Example      &\#pred&size dom&size pre&\#pre&size int& \#int/pre \\
\hline			                                  
appendlast	& 2 &    3 &    12 &$3^{12}$&   13 & $2^{13}$ \\
reverselast	& 2 &    3 &    12 &$3^{12}$&   13 & $2^{13}$ \\
nreverselast	& 3 &    5 &    28 &$5^{28}$&  150 &$2^{150}$ \\
\hline				    	              
multiset1o	& 1 &    2 &     7 & $2^{7}$&    4 &  $2^{4}$ \\
multiset2o	& 1 &    2 &     7 & $2^{7}$&    4 &  $2^{4}$ \\
multiset3o	& 1 &    2 &     7 & $2^{7}$&    4 &  $2^{4}$ \\
\hline				    	           	   
blockpair2o	& 3 &    2 &    19 &$2^{19}$&   12 & $2^{12}$ \\
blockpair3o	& 3 &    2 &    36 &$2^{36}$&   20 & $2^{20}$ \\
blockpair2l	& 5 &    2 &    19 &$2^{19}$&   32 & $2^{32}$ \\
blockpair3l	& 5 &    2 &    36 &$2^{36}$&   40 & $2^{40}$ \\
blockzero2o	& 3 &    2 &    19 &$2^{19}$&   12 & $2^{12}$ \\
blockzero3o	& 3 &    2 &    35 &$2^{35}$&   20 & $2^{20}$ \\
blockzero2l	& 5 &    2 &    19 &$2^{19}$&   32 & $2^{32}$ \\
blockzero3l	& 5 &    2 &    35 &$2^{35}$&   40 & $2^{40}$ \\
blockzero2ls	& 5 &    2 &    19 &$2^{19}$&   32 & $2^{32}$ \\
\hline				    	          	
tba (BOO019-1)	& 1 &    3 &    32 &$3^{32}$&    9 &      1   \\
grp (GRP081-1)	& 1 &    2 &    17 &$2^{17}$&    4 &      1   \\
cl3 (COL005-1)	& 1 &    3 &    12 &$3^{12}$&    9 &      1   \\
\hline
\end{tabular}
\end{center}
\end{table}

\subsection{Results} \label{subsec:res}

The results with \FMC were taken from \cite{mg:peltier-98} or were
sent to us by its author which was using a SUN 4 ELC machine. 
All other systems were run on SUN Sparc Ultra-2 computer. 
The system AB is the abductive system described in \cite{fq:mb-jflp},
however, running under (the slower) XSB-Prolog instead of Master
Prolog for equal comparison.
We used FINDER \cite{mg:finder} version 3.0.2 and SEM \cite{mg:sem}
version 1.7 which are well known model generators implemented in C.

The system {\em naive} results from the direct translation of the
system AB to XSB: it uses the same failure analysis, it starts from a
random total order over the cells of the pre-interpretation and it
uses the simplest variant of {\tt check\_return} which sticks
to the first answer whatever the associated conflict set is. 
For the TPTP problems the standard equality axioms were used.

The systems {\em single~CS} and {\em best~CS} use a more sophisticated
version of {\tt check\_return} which prefers the answer with the
shorter conflict set, advanced failure analysis and the more
sophisticated version of intelligent backtracking which leaves
elements unordered until they participate in a conflict set.  The
system {\em single~CS} uses the first answer to the top level query to
direct the backtracking. The system {\em best~CS} computes all answers
to the top level query and then selects from them the conflict set
which will add the fewest number of cells to the ordered sequence.
Both systems use the technique described in Section~\ref{sec:eq} on
the TPTP problems.

Table~\ref{tab:times} gives the times obtained by the different systems.
The time is in seconds unless followed by H, then it is in hours.
A ``-'' means the example was not run. A ``$>n$'' means the system had
still no solution after time $n$.

\begin{table}
\caption{Execution times} \label{tab:times}
\begin{center}
\begin{tabular}{|l|r|r|r|r|r|r|r|}
\hline
Example	      & naive &single CS& best CS&    AB  & FINDER &   SEM & \FMC   \\
\hline
appendlast	&   919 &  0.76 &   1.63 &   1.42 &   0.07 &  0.01 &  45.21 \\
reverselast	&   918 &  0.85 &   1.85 &   1.00 &   0.10 &  0.01 &  10.79 \\
nreverselast    &$>$2706&$>$1673&    178 &  17.5H &$>1446$ &   957 & $>$900 \\
\hline		         	  	   	                            
multiseto1	&  0.18 &  0.06 &   0.12 &   0.08 &   0.02 &  0.01 &      - \\
multiseto2	&  0.07 &  0.20 &   0.47 &   0.10 &   0.02 &  0.01 &   0.02 \\
multiseto3	&  0.94 &  0.54 &   2.77 &   0.28 &   0.03 &  0.01 &      - \\
\hline		         	  	   	                            
blockpair2o	&   451 &  0.86 &   3.14 &   5.05 &   0.07 &  0.05 &   7.31 \\
blockpair3o	& $>$58 &  0.94 &   3.90 &  21.97 &   0.18 &  0.23 & $>$900 \\
blockpair2l	&  5303 &  1.86 &   7.85 &   3.56 &   0.04 &  0.05 &  204.9 \\
blockpair3l	&$>$222 &  2.05 &   9.70 &  53.88 &   0.12 &  0.18 & $>$900 \\
blockzero2o	&  7.93 &  7.94 &   4.35 &   2.84 &   0.11 &  0.09 &      - \\
blockzero3o	&   162 &  8.86 &   5.41 &  24.48 &   0.22 &  1.98 &      - \\
blockzero2l	& 18.49 &  2.00 &  20.71 &   5.67 &   0.23 &  0.10 &      - \\
blockzero3l	& 40.35 &  2.06 &  24.76 &  37.23 &   0.33 &  2.39 &      - \\
blockzero2ls    & 11.8H &   648 &   2631 &    593 &   2287 &  5.05 & $>$900 \\
\hline		         	  	   	                            
tba		&$>$950 &  1331 &   3.65 &   3.29 &   0.03 &  0.03 &   0.06 \\
grp		&  1189 &  1.05 &   5.89 &  13.94 &   0.03 &  0.01 &      - \\
cl3		&  0.13 &  3.85 &   1.63 &   1.03 &   0.02 &  0.03 &   0.04 \\
\hline
\end{tabular}
\end{center}
\end{table}

Table~\ref{tab:back} shows the number of generated and tested
pre-interpretations (number of backtracks). 
For the SEM system, we have modified the source code to report exactly
this number.
For the FINDER system we report the sum of the number of 
{\em bad candidates tested} and {\em other backtracks}. 
Also in this table ``-'' means not run, ``$>n$'' means already $n$
backtracks when interrupted. 
For the system {\em best~CS} we give an additional column {\em total} 
which shows the total number of conflict sets obtained as ``answers''
to the query (divided by the number of backtracks, this gives the
average number of conflict sets obtained when running the query).

\begin{table}
\caption{Number of backtracks} \label{tab:back}
\begin{center}
\begin{tabular}{|l|r|r|r|r|r|r|r|r|}
\hline
Example	 	& naive  & single CS &
               \multicolumn{2}{c|}{best CS}   &  AB  & FINDER&  SEM  & \FMC   \\
\cline{2-9}
                &\#bckt  &\#bckt&\#bckt& total&\#bckt& \#bckt& \#bckt& \#bckt \\
\hline		
appendlast	&  41045 &   56 &   27 &  136 &   43 &   180 &    27 & 110019 \\
reverselast	&  41045 &   56 &   27 &  133 &   30 &   211 &    27 &  23445 \\
nreverselast    &$>$10000&$>$2000& 221 & 2426 &190170&$>10^7$&31285086&  $>$? \\
\hline		         				      
multiset1o	&      4 &    3 &    3 &   11 &    4 &     4 &     3 &      - \\
multiset2o	&     14 &   14 &   12 &   38 &   10 &	  31 &     8 &    104 \\
multiset3o	&    127 &   75 &   76 &  122 &   33 &	  75 &    86 &      - \\
\hline		         
blockpair2o	&   9323 &   34 &   32 &   55 &   17 &   273 &   918 &   5567 \\
blockpair3o	&$>$3000 &   34 &   32 &   55 &   56 &   879 &  2904 &   $>$? \\
blockpair2l	&  32873 &   76 &   66 &  117 &   33 &    68 &   918 &  91404 \\
blockpair3l	&$>$6000 &   76 &   66 &  117 &  204 &   359 &  2904 &   $>$? \\
blockzero2o     &    577 &  241 &   48 &  148 &  158 &   823 &  3495 &      - \\
blockzero3o	&   1245 &  241 &   48 &  148 &  500 &   897 & 63032 &      - \\
blockzero2l	&   1145 &  190 &  181 & 1044 &   98 &  1131 &  3415 &      - \\
blockzero3l	&   2289 &  190 &  181 & 1044 &  380 &  1123 & 63288 &      - \\
blockzero2ls	& 128926 &21544 &20284 &31969 & 3615 &3999226&201882 &      - \\
\hline		         
tba		&$>$4000 &95369 &   41 &   91 &   72 &	  23 &     5 &     33 \\
grp		&  19996 &   71 &  138 &  210 &  361 &	  24 &    14 &      - \\
cl3		&      5 &  670 &   93 &  191 &   41 &	  30 &     3 &      - \\
\hline		        
\end{tabular}
\end{center}
\end{table}


\subsection{Discussion}

Comparing the systems {\em naive} and AB, we see that the
straightforward transfer of AB to XSB results in a much worse
behavior. Hence the heuristics used by AB to control the search have
a big impact.

The effect of the advanced failure analysis is not reported
separately. Its impact is only visible in the {\tt block*3?} problems
which compute, for the failure analysis, an irrelevant output argument.
The advanced failure analysis makes these problems behave as well as
the corresponding {\tt block*2?} problems.  Note that the AB system as
well as all first order model generators behave much worse on the
3-argument problems than on the corresponding 2-argument problems.
As computing some output is a natural feature of a logic program, the
advanced failure analysis is an important asset of our system.

Adding more sophisticated backtracking which does not fix the order of
the cells in advance yields a substantial improvement on most
problems. The system {\em single~CS} which sticks everywhere to the
first conflict set is often the fastest, although it often needs more
backtracks than {\em best~CS}. It fails only on {\tt nreverselast}
which uses a 5 element domain and has a very large search space.
However, on the equality problems it becomes obvious that a good
choice of a conflict set is essential for solving such problems.
In number of backtracks, {\em best~CS} compares quite well with
AB. Only on {\tt blockzero2ls} it needs a lot more backtracks,
while it needs a lot less on {\tt nreverselast}. Perhaps on 
{\tt blockzero2ls}, which has no solution, it suffers from the less
optimal ordering because the search space has to be searched
exhaustively.

From the model generators FINDER and SEM perform reasonably well in
terms of time and also in number of backtracks. However, the results
for FINDER were obtained only after a fine tuning of the different
parameters and the representation of the problems (see
\cite{fq:mb-jflp}). The system also uses intelligent backtracking for
deriving secondary conflict sets and some other forms of failure
analysis. It has a smaller number of backtracks on the more complex
planning problems than SEM.  The system SEM is the fastest in raw
speed and is not so sensible to the problem representation.
Of the model generators, the system \FMC is the weakest on the class
of problems we consider. This result contrasts with the results in
\cite{mg:peltier-98} where it is the best on several problems.

Compared with our system the model generators have to backtrack much
more on the planning problems and the other logic programs where they
have to explore the full space of interpretations while we look only
for the least model of the program for a given pre-interpretation (the
extra cost of evaluating the query in the least model is more than
compensated for by the exponentially smaller search space). On the
TPTP problems our system is doing worse which suggests that there is
further room for making better use of the information in conflict
sets.

\section{Conclusion} \label{sec:concl}

In this paper we presented a method for proving failure of queries for
definite logic programs based on direct execution of the abstracted
program in XSB-Prolog, a standard top-down proof procedure with tabulation. 

By using a better form of intelligent backtracking (proposed in
\cite{ib:mb-81}) which does not fix the enumeration order in
advance and an improved failure analysis, we were able to compensate
for the loss of flexibility which results from the direct execution of
the abstracted program.

This way of intelligent backtracking could also be interesting for
other systems, e.g.\ \FMC of which Peltier reports that it is quite
sensitive to the initial enumeration order.

While difference in speed with the AB system are modest, the approach
is still very interesting as the depth-first left-to-right execution
results in a much better memory management so that larger problems can
be tackled. The meta-interpreter of the AB system keeps track of the
whole top-down proof tree in evaluating the query, which leads to very
large memory consumption.

Interesting future work is to further investigate some control issues.
One could explore whether there is a good compromise between computing
only one solution to the query and computing all solutions. One could
try to further improve the backtracking by developing some heuristics
which order a group of new elements when they are inserted in the
ordered sequence.

\section*{Acknowledgements}

We want to thank Kostis Sagonas for his help with the XSB system.
Maurice Bruynooghe is supported by FWO-Vlaanderen. Nikolay Pelov is
supported by the GOA project LP+.

\appendix

\section{Code for Some of the Problems} \label{app:code}

\subsection{Multiset}
\begin{verbatim}
multiset1o :- sameMultiSet(a, X), sameMultiSet(X, b).
multiset2o :- sameMultiSet(o(a,o(a,emptyMultiSet)),o(X,o(emptyMultiSet,b))).
multiset3o :- sameMultiSet(o(a,o(a,o(emptyMultiSet,b))), 
                   o(o(a,b),o(a,emptyMultiSet))).

sameMultiSet(X, X).
sameMultiSet(o(X, Y), o(X, Z)):- sameMultiSet(Y, Z).
sameMultiSet(o(o(X, Y), Z), U):- sameMultiSet(o(X, o(Y, Z)), U).
sameMultiSet(U, o(o(X, Y), Z)):- sameMultiSet(U, o(X, o(Y, Z))).
sameMultiSet(o(emptyMultiSet, X), Y):- 	sameMultiSet(X, Y).
sameMultiSet(X, o(emptyMultiSet, Y)):-	sameMultiSet(X, Y).
sameMultiSet(o(X, Y), Z) :- sameMultiSet(o(Y, X), Z).
\end{verbatim}

\subsection{Planning Problems}

Blocks are identified by integers represented as terms with the
constant $0$ and the function $s/1$. The $actionZero/3$ predicate
gives the possible actions and the $causesZero/3$ predicate tries to
find a plan. In both predicates the first argument is the initial
state, the last argument is the final state and the plan is collected
in the second argument.

\begin{verbatim}
blockzero3o :-
    causesZero(o(o(on(s(s(0)), s(0)), cl(s(s(0)))), em), Plan,
               o(on(s(0), 0), Z)).

causesZero(I1, void, I2):-
    sameMultiSet(I1, I2).
causesZero(I, plan(A, P), G):-  
    actionZero(C, A, E),
    sameMultiSet(o(C, Z), I),
    causesZero(o(E, Z), P, G).

actionZero(holds(V), put_down(V), 
           o(table(V), o(clear(V), nul))).
actionZero(o(clear(V), o(table(V), nul)), pick_up(V), 
           holds(V)).
actionZero(o(holds(V), clear(W)), stack(V, W), 
           o(on(V,W), o(clear(V), nul))).
actionZero(o(clear(V), o(on(V, W), nul)), unstack(V), 
           o(holds(V), clear(W))).
actionZero(o(on(X, Y), o(clear(X), nul)), generate_block,
           o(on(s(X), X), o(on(X, Y), o(clear(s(X)), nul)))).
\end{verbatim}



\end{document}